# Diamond spin sensors: A new way to probe nanomagnetism

Jean-Philippe TETIENNE[1], Liam P. McGUINNESS[1,2] and Vincent JACQUES[3]

[1]Centre for Quantum Computation and Communication Technology, School of Physics, University of Melbourne, Parkville, Victoria 3010, Australia
[2]Institute for Quantum Optics and Center for Integrated Quantum Science and Technology, University of Ulm, 89081 Ulm, Germany
[3]Laboratoire Charles Coulomb, Université de Montpellier and CNRS, 34095 Montpellier, France

# 1. Introduction

Studies of individual quantum systems, which have led to considerable progress in our understanding of quantum physics, have traditionally been associated with atomic gases. In the last decades however, the emphasis has shifted towards solid-state systems, which are much more practical for applications. In particular, a new field has recently emerged that is concerned with the study of quantum systems based on single spins localized near point defects in crystalline solids. One such system is the nitrogen-vacancy (NV) defect in diamond. Initially used as an experimental breadboard for testing concepts of quantum physics and quantum computation, the NV defect was soon proposed as a sensitive magnetometer, capable of detecting minute magnetic fields, down to ultimate level of single spins. This atomic-sized magnetometer can be used as a standalone sensor, or integrated into an imaging system providing spatial resolution down to the atomic scale. Diamond-based instruments thus offer new pathways to probe the magnetism of matter from the mesoscale down to the nanoscale.

This chapter gives an overview of the field of diamond-based magnetic sensing and imaging, with an emphasis on already demonstrated applications of this technology. The chapter is divided into three main sections. In Section 2, the underlying physics and methods of diamond-based magnetometry are described. Section 3 is devoted to various experimental implementations that employ this new class of sensors for magnetic sensing and imaging. Finally, some recent applications are presented in Section 4.

# 2. Magnetic sensing with nitrogen-vacancy defects in diamond

The NV defect in diamond consists of a substitutional nitrogen atom adjacent to a vacancy, *i.e.* a missing carbon atom (Fig. 1a) [Doherty2013]. Although this defect has been known since the 1960's from spectroscopic measurements [Zaitsev2001,Davies1974,VanOort1988], it was only in 1997 that it could be detected at the level of individual sites, using a confocal microscope operating under ambient condition [Gruber1997]. Nearly ten years later, it was proposed that a single NV defect can act as a sensitive, atomic-sized magnetometer [Chernobrod2005,Taylor2008,Degen2008]. The basic principle relies on microwave-induced electron spin transitions of the NV defect, which are detected via a change of its photoluminescence. This method allows Zeeman shifts of the electron spin resonance to be monitored, hence providing a direct access to the magnitude of the magnetic field impingent on the diamond. Following the first experimental demonstrations based on this idea [Maze2008,Balasubramanian2008], a range of new sensing schemes have been proposed and experimentally tested, in particular to detect randomly fluctuating magnetic fields emanating, *e.g.*, from paramagnetic molecules [Cole2009,Hall2009].

In the following, we will first provide a basic introduction to the physics of the NV defect and its main properties (§2.1). We will then describe the main strategies that are used to exploit these properties in order to measure magnetic fields (§2.2).

## 2.1 Physics of the NV defect in diamond

Like many point defects in semiconductors, the NV defect gives rise to localized electronic states [Gali2008,Larsson2008,Hossain2008]. Because their energies lie deeply within the band gap of diamond, these states are essentially decoupled from the valence and conduction bands.

As a consequence, the NV defect behaves very much like a molecule "*nestled*" in the diamond matrix. It is generally found in its negative charge state NV⁻, which means that an additional electron is donated by a nearby impurity, leading to two unpaired electrons at the defect location. In the following, the negative charge state NV⁻ will be simply referred to as the NV defect.

*Optical properties*. The NV defect exhibits strong optical transitions as shown in Fig. 1b, which allows individual defects to be observed with confocal microscopy [Gruber1997]. It is efficiently excited by a green laser, typically at a wavelength λ = 532 nm, and relaxes towards the ground state by emitting Stokes-shifted red photoluminescence (PL) at λ ≈ 650-750 nm. Importantly, the NV defect is highly photostable at room temperature. This allows reliable assignment of small fluorescence variations to physical properties of the defect, *e.g.,* a change of spin state.

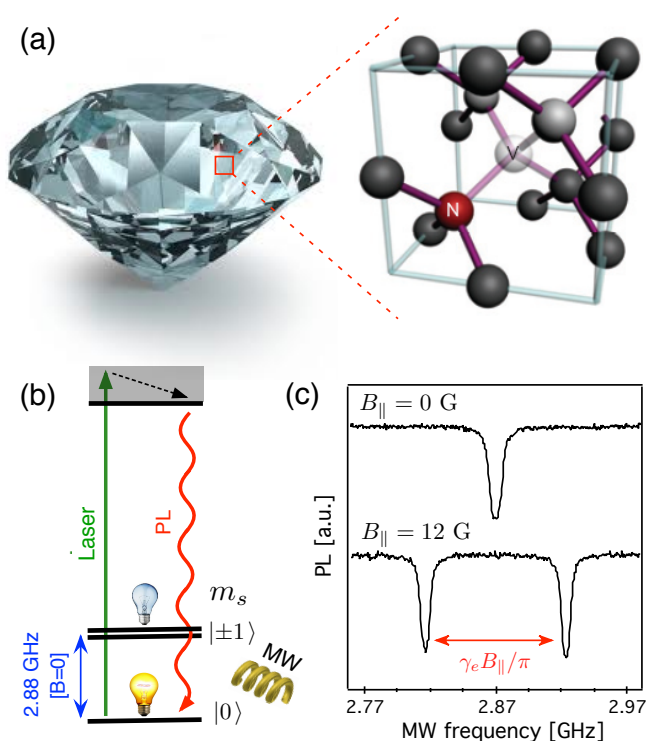

**Figure 1**: *(a) Atomic structure of the NV defect in diamond. (b) Simplified energy level scheme. The NV defect is polarized into the spin sublevel |0⟩ by optical pumping, and exhibits a spin-dependent photoluminescence (PL) intensity. (c) Optically detected magnetic resonance (ODMR) spectra recorded for different magnetic field magnitudes $B_\parallel$ applied to a single NV defect in diamond. The ODMR transitions are shifted owing to Zeeman effect thus providing a measurement of the magnetic field projection $B_\parallel$ along the NV defect quantization axis. These spectra are recorded by monitoring the NV defect PL intensity while sweeping the frequency of the microwave field.*

*Optically detected magnetic resonance.* The electronic ground state of the NV⁻ defect is a spin triplet $S = 1$, with a fine structure resulting from the interaction between the two unpaired electrons hosted by the NV defect [Doherty2011,Maze2011]. More precisely, spin-spin interaction induces a splitting of ≈ 2.87 GHz between the spin sublevel |0⟩ and the doublet |±1⟩ (Fig. 1b). Here $|m_s⟩$ denotes the electronic spin state, with the number $m_s$ being the spin projection along the symmetry axis of the NV defect. A key property of the NV defect is the ability to initialize and read out its spin state by optical means. Indeed, spin-dependent non-radiative decay through metastable singlet states are responsible for the two following effects: (i) under continuous optical excitation, the NV defect's spin is polarized into the |0⟩ state with high fidelity, ≥ 90%; (ii) the |0⟩ state exhibits a larger PL rate than the |±1⟩ states [Manson2006,Robledo2011]. These two properties enable optical detection of the electron spin resonance of a single NV defect, a technique known as "optically detected magnetic resonance" (ODMR). Experimentally, this is typically performed by measuring the PL intensity as a function of the frequency of an applied microwave (MW) field. As shown in Fig. 1c, a drop of the PL signal is observed when the MW frequency matches the spin transitions $|0\rangle \rightarrow |\pm 1\rangle$.

***Magnetometry***. In the presence of a magnetic field **B**, the degeneracy between the states $|\pm 1\rangle$ is lifted by the Zeeman interaction, leading to the appearance of two resonance lines in the ODMR spectrum (Fig. 1c), whose splitting is directly linked to the amplitude of the local magnetic field [Rondin2014]. Furthermore, quantum coherent control of the NV spin state can be achieved by applying short microwave pulses. This allows preparation of quantum superpositions, which are very sensitive to magnetic field fluctuations and hence provide a way to probe them. In the next section, the main techniques developed to measure static and fluctuating magnetic fields with high sensitivity are described.

## 2.2 Magnetic sensing methods

In this section, we outline the main magnetometry techniques relying on the NV defect in diamond. We begin with ODMR spectroscopy, which enables measurement of slowly varying or static (DC) magnetic fields. We then present techniques to detect time-varying magnetic fields – using phase measurements to detect AC fields or randomly fluctuating "magnetic noise" in the kilohertz to megahertz frequency range; or using spin relaxometry to probe magnetic noise in the gigahertz range. The sensitivity of each approach to magnetic fields is also discussed.

***ODMR spectroscopy.*** The most straightforward way to measure the static component of the local magnetic field using an NV defect is to monitor the ODMR spectrum, i.e., the PL intensity measured as a function of the MW frequency under continuous laser and MW excitation (Fig. 2a). For small magnetic field amplitudes (typically < 10 mT), the Zeeman splitting between the transitions $|0\rangle \rightarrow |+1\rangle$ and $|0\rangle \rightarrow |-1\rangle$ is given by $\Delta \nu = \gamma_e B_\parallel / \pi$, where $\gamma_e$ is the electron gyromagnetic ratio and $B_\parallel$ is the projection of the magnetic field along the NV defect's symmetry axis [Rondin2014]. Thus, ODMR spectroscopy provides a direct measure of the field component $B_\parallel$. For larger amplitudes (> 10 mT), the Zeeman shift also depends on the transverse component of the magnetic field $B_\perp$, allowing the magnetic field orientation to be determined, however this is accompanied by a reduction in optical contrast, and hence sensitivity with increasing $B_\perp$ [Tetienne2012].

The magnetic sensitivity of the method is governed by the photon shot noise of the PL signal, which limits the precision to which the Zeeman shift $\Delta \nu$ can be determined. The sensitivity is conventionally characterized by the quantity $\eta = \delta B \sqrt{\Delta t}$, where $\delta B$ is the smallest change in the magnetic field $B_\parallel$ detectable with a signal-to-noise ratio of 1 and an integration time $\Delta t$. It can be shown that the optimum sensitivity scales as $\eta \propto 1/\sqrt{T_2^*}$, where $T_2^*$ is the inhomogeneous spin dephasing time which limits the line-width in the ODMR spectrum [Taylor2008, Dreau2011]. At room-temperature $T_2^*$ varies greatly, from less than 100 ns to more than 100 μs, depending on the host diamond [Zhao2012]. Under typical experimental conditions, this leads to sensitivities between 0.1 and 10 μT/$\sqrt{\text{Hz}}$ for a single NV defect.

***Spin phase sensing.*** While ODMR spectroscopy is inherently slow and usually limited to sub-kHz variations of the magnetic field, fluctuations in the kHz-MHz range can be probed using spin phase measurements [Maze2008,Cole2009,Hall2009,Meriles2010]. The basic idea is to measure the coherent phase $\delta \phi$ acquired by the NV spin initially prepared in a quantum superposition $|\psi\rangle = \frac{1}{\sqrt{2}}(|0\rangle + |1\rangle)$. The spin state evolves according to $|\psi(\tau)\rangle = \frac{1}{\sqrt{2}}(|0\rangle +$

$e^{-i\delta\phi(\tau)}|1\rangle)$ and the phase $\delta\phi(\tau)$ after a time $\tau$ is then converted into a population difference $\cos\delta\phi$ which is read out optically. This is typically achieved by means of a spin echo sequence with a total evolution time $\tau$ (Fig. 2b). In the presence of an oscillating (AC) magnetic field of frequency $\nu = 1/\tau$ and amplitude $B_{ac}$, the resulting phase $\delta\phi$ is directly proportional to $B_{ac}$, hence phase measurements provide a way to determine the amplitude of a field of known frequency, as illustrated in Fig. 2b. Furthermore, echo-based sequences allow the frequency of the external field to be determined, by adjustment of the evolution time. Compared with DC magnetometry, the optimum sensitivity of AC magnetometry is enhanced by a factor $\sqrt{T_2/T_2^*}$, where $T_2$ is the homogenous spin dephasing time and can exceed 1 ms at room temperature [Balasubramanian2009]. This leads to sensitivities of the order of $10 \text{ nT}/\sqrt{\text{Hz}}$ for a single NV defect.

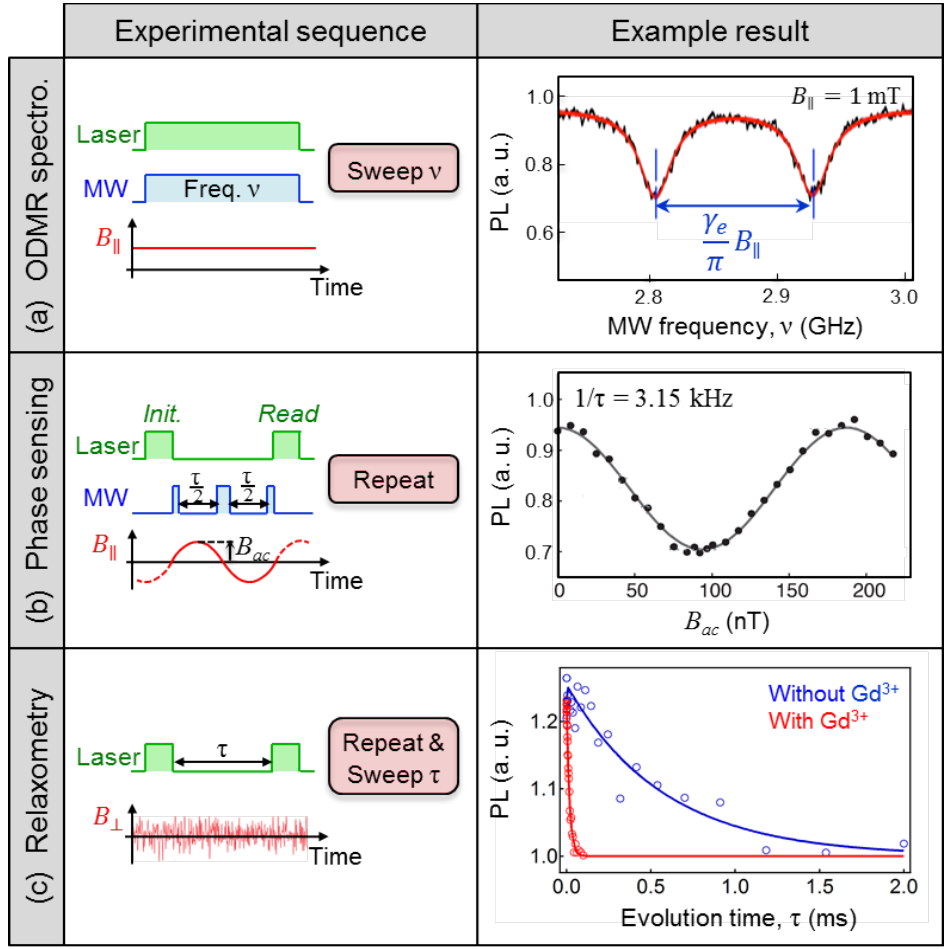

**Figure 2**: *Illustration of the basic magnetometry techniques based on the NV defect in diamond: (a) ODMR spectroscopy to measure DC magnetic fields, (b) spin phase sensing to measure kHz-MHz AC magnetic fields or magnetic noise and (c) spin relaxometry to measure GHz magnetic noise (c). For each case, the left-hand column shows the sequence of laser (green) and MW (blue) pulses applied to perform the measurement, while the right-hand column shows a typical experimental result. The graph in (b) is adapted from [Maze2008].*

For a randomly fluctuating magnetic field, *i.e.* magnetic noise, the acquired phase cancels out on average but nevertheless the finite variance $\langle\delta\phi(\tau)^2\rangle$ induces a decrease in population hence in PL signal after a spin echo sequence. In general, this translates into a monotonic decay of the PL signal against evolution time $\tau$, where the characteristic decay time defines the spin dephasing time $T_2$. This time is governed by the strength of the magnetic noise at frequencies in the kHz-MHz range. This provides a way to probe the presence of paramagnetic species within the diamond but also outside [Hall2009,Meriles2010], such as magnetic ions in a liquid solution [McGuinness2013] or attached to the diamond surface [Ermakova2013]. The sensitivity of the technique is similar to that of AC magnetometry, limited by the intrinsic $T_2$ time. While dephasing in the spin echo sequence is sensitive to a broad range of frequencies, narrower spectral filtering can be achieved by applying trains of MW pulses [deLange2011]. This enables measurement of the noise spectrum of the local magnetic field, as discussed in §4.2.

***Spin relaxometry.*** Higher frequencies – in the GHz range – are more conveniently probed using population relaxation rather than dephasing [Steinert2013,Ermakova2013, Tetienne2013,Kaufmann2013,Ziem2013,Pelliccione2014]. In this approach, laser pulses are used to prepare the spin into the |0⟩ state and observe its return to equilibrium, governed by the spin relaxation time $T_1$ (Fig. 2c). The $T_1$ time is sensitive to magnetic noise at the spin transition frequencies, which are around 3 GHz for small background fields. As an illustration, Fig. 2c shows relaxation curves measured from an NV defect in a nanodiamond: the blue curve corresponds to the bare nanodiamond in air, indicating an intrinsic $T_1 \approx 600~\mu s$, while the red curve shows a strong reduction ($T_1 \approx 20~\mu s$) after decorating the nanodiamond with paramagnetic $Gd^{3+}$ molecules. It was recently shown that this technique can reach the single molecule detection level [Sushkov2014] and also be used to unravel Johnson noise and ballistic transport of electrons in metallic films [Kolkowitz2015].

## 3. Experimental implementations for sensing and imaging

The NV defect derives much of its value as a nanoscale sensor from the ability to *optically* readout magnetic fields at a distance. This property sets it apart from other solid-state magnetometers such as Hall probes, magnetic force microscopes and SQUIDs, since no electrical contact, or physical contact of any sort is required with the sensor. This reduction in components allows for miniaturization, and is a key benefit in the non-invasiveness and nanoscale resolution of the NV system. The read-out of magnetic fields is performed simply by collecting the PL emitted from the NV center, based on the principles discussed in the previous section. However, in order to benefit from its small size and address nanometric objects, the NV center probe must be approached as close to the sample as possible. This is because the probe-to-sample distance usually limits the spatial resolution of the measurement, that is, the degree of spatial detail one can infer about the sample. In addition, the probe-to-sample distance affects the strength of the signal to be detected; i.e. the larger the stand-off distance, the weaker the signal.

In this section, we present the main experimental configurations that are implemented to bring NV center probes in the vicinity of magnetically active samples – down to a few nanometers – and form magnetic images. We distinguish three different approaches, which employ: a single scanning NV defect (§3.1), a single stationary NV defect (§3.2), or wide-field detection of an ensemble of NV defects (§3.3). Finally, we will discuss the challenges and future improvements towards optimal device construction (§3.4).

## 3.1 With a scanning NV defect

Perhaps the most obvious way to bring an NV magnetometer close to a sample is to position the sensor onto a nano-manipulator, such as the tip of an atomic force microscope (AFM) cantilever. Practically, this can be achieved by grafting a nanometer-sized diamond, a nanodiamond, containing one or several NV defects onto the tip apex [Balasubramanian2008, Rondin2012,Tetienne2016] (Fig. 3a). By integrating the AFM with an optical confocal microscope to enable laser excitation and readout of the NV defect PL, it is possible to scan the NV sensor relative to the sample with nanometer precision, and form images of the stray magnetic field produced by the sample by recording ODMR spectra at each pixel of the scan.

A typical experimental result provided by this approach is shown in Fig. 3 [Tetienne2013b]. Here, the sample is a thin 1-μm-diameter disc made of $Fe_{20}Ni_{80}$, which is a soft ferromagnetic alloy (see AFM image in Fig. 3b). In such micrometer-sized dots, the equilibrium magnetic state is a vortex state, that is, the magnetization lies in the disc's plane and curls about the center. At the disc's center, the magnetization points out of the plane within a region of typically ~10 nm in size, called the vortex core. The only source of stray magnetic field outside the magnetic structure arises from the vortex core, which acts as an effective point-like magnetic dipole oriented perpendicular to the surface. Fig. 3c shows the full magnetic field distribution above the vortex core recorded with the scanning-NV magnetometer (projected along the NV symmetry axis). Here the width of the magnetic field spot is given by the probe-to-sample distance, which is about 100 nm in this experiment. A shorter distance would give a smaller spot – i.e., a better spatial resolution – down to the limit where the spot size would reflect the actual size of the vortex core. This illustrates a key challenge of scanning-NV magnetometry, which is the reduction of the probe-to-sample distance.

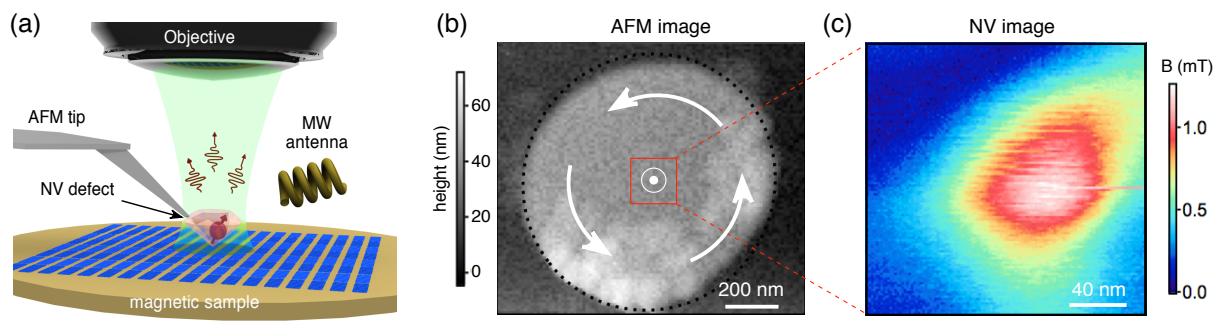

**Figure 3**: *(a) Principle of scanning-NV magnetometry. (b) AFM image of a $Fe_{20}Ni_{80}$ circular ferromagnetic dot (1 μm diameter). The white arrows indicate the in-plane curling magnetization of the vortex state. At the center of the structure, the magnetization is out-of-plane with a typical length scale of 10 nm. (b) Full magnetic field distribution above the vortex core recorded with the scanning-NV magnetometer. The width of the magnetic field distribution is limited by the probe-to-sample distance (~100 nm). Adapted from [Rondin2012] and [Tetienne2013b].*

In the scanning-NV arrangement, critical factors for imaging resolution are *i)* to ensure the nanodiamond is placed directly at the tip apex, and *ii)* to reduce the size of the nanodiamond in order to have the NV defect placed as close to the sample as possible. Through optimization of these factors, a distance as small as ~16 nm has been achieved [Tisler2013]. However, there is a fundamental limitation when using nanodiamonds. Indeed, the magnetic sensitivity depends primarily on the diamond quality as this determines the NV coherence time (see §3.4). Currently, the quality of nanodiamond material is inferior to that of bulk diamond, meaning the magnetic sensitivity is generally limited to a few $\mu T/Hz^{1/2}$. A possible route to realize high-sensitivity scanning-NV magnetometers is to integrate a single NV defect into an all-diamond AFM tip fabricated from a high-quality bulk diamond crystal [Maletinsky2012]. Notably, this strategy has enabled the detection and imaging of the stray magnetic field emanating from a single electron spin at room temperature [Grinolds2013].

## 3.2 With a stationary NV defect

One drawback associated with scanning-NV magnetometry is the need to place the NV center on a scanning tip while minimizing the NV-to-sample distance. By placing the sample rather than the diamond on a scanning tip, strict requirements on the geometry of the diamond sensor can be relaxed. In this configuration a large diamond sample containing NV centers close to a flat surface is used and the sample is scanned over an underlying NV center (Fig. 4a) [Pelliccione2014,Häberle2015,Rugar2015,Schmid-Lorch2015]. A further benefit of this geometry is that extra care can be taken to grow extremely high purity diamond whilst avoiding any potential residual damage that may result from fabrication of the diamond into tips or using nanodiamonds.

Nevertheless, scanning sample configurations require that the sample of interest can be attached to a scanning tip, which hampers studies of bulk samples or materials that require electrical contacts. Apart from an inversion of the sensor and sample, the detection techniques for stationary NV magnetometry remain the same as scanning NV magnetometry, and comparable resolution and sensitivity can be achieved. As an illustration, Fig. 4 describes an experiment where the sample is an organic micro-sphere attached to an AFM cantilever [Häberle2015]. When the tip is engaged, the fluctuating magnetic field produced by the nuclear spins in the sample is detected by the NV center placed ~5 nm below the diamond surface. Using advanced protocols (see §4.2), it is even possible to selectively detect specific nuclear species ($^{19}$F in this example). Scanning of the sample (here a grating engraved in the microsphere, see Fig. 4a) relative to the NV center then enables the formation of images with chemical contrast, as shown in Fig. 4b.

Of note, near-surface NV defects can also be used to investigate samples that are prepared directly on the diamond surface, without incorporating a scanning device. This configuration is particularly suited for dynamic studies where spatial information is not required; although several NV defects may probe different regions of the sample. For instance, by coating samples on the diamond surface, the dynamics of single spin-labeled proteins [Shi2015], ferritin molecules [SchäferNolte2014], as well as spin waves in ferromagnetic micro-discs [vanderSar2015] and Johnson noise in metallic films [Kolkowitz2015], have been investigated. Furthermore, imaging can be achieved with this sample-on-diamond geometry in instances where the NV-sample interaction can be spatially modulated. In particular, magnetic field gradients can be applied to scan the "resonance slice" in spin ensembles in a similar fashion as in magnetic resonance imaging (MRI) [Grinolds2014]. This, in complement with Fourier deconvolution techniques [Arai2015,Lazariev2015], could lead to atomic-resolution imaging of individual small molecules.

Finally, it is worth mentioning another use of NV centers as standalone sensors. When integrated in nanodiamonds, they can be deployed inside living cells [McGuinness2011]. In effect, a nanodiamond becomes a molecular-sized reporter of magnetic field, which can be read-out using optical microscopy under physiological conditions, with unique applications for biosensing.

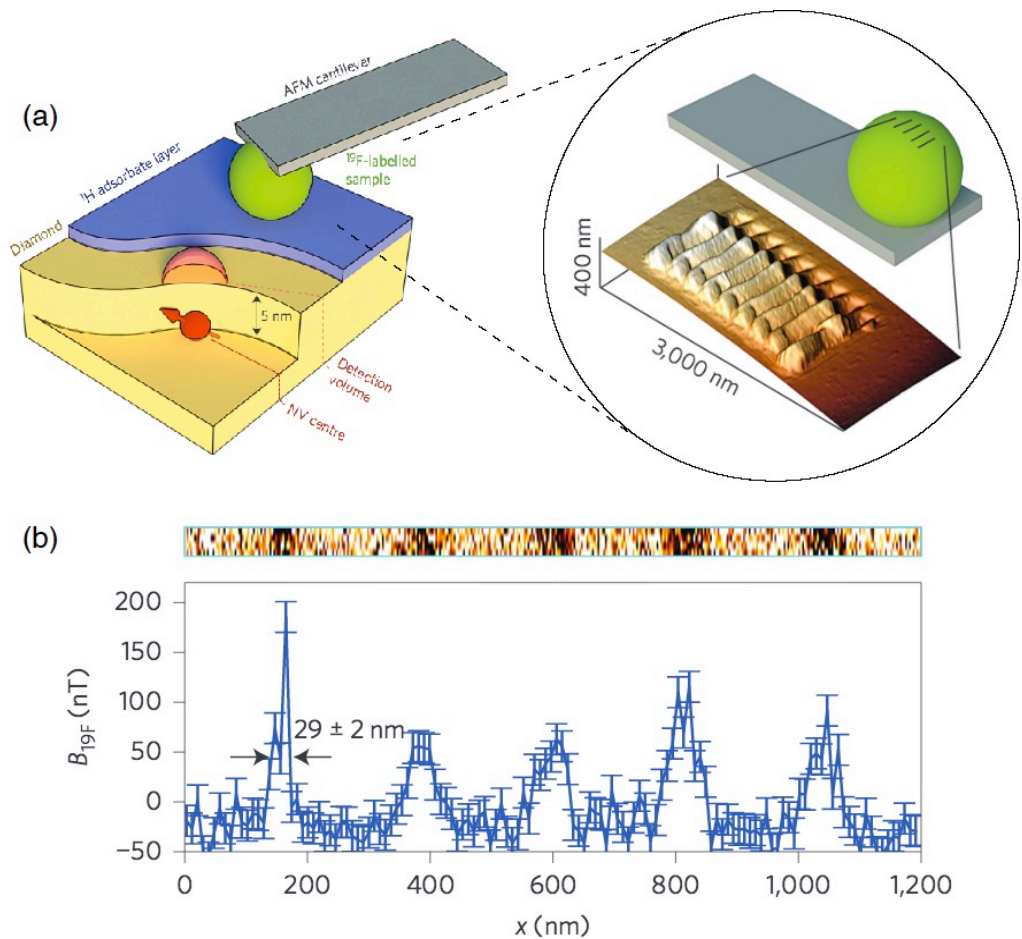

**Figure 4**: *(a) Principle of scanning sample magnetometry with stationary NV. Here the sample is a $^{19}$F-rich organic micro-sphere. The inset shows an AFM image of the grating engraved into the sphere. (b) $^{19}$F-resolved image of the sample (top). Dark regions indicate a large $^{19}$F signal. The graph (bottom) is a profile obtained by binning the image. The smallest resolvable feature of 29 nm is limited simultaneously by the NV-to-sample distance and the finite sharpness of the grating. Adapted from [Häberle2015].*

### 3.3 Wide-field imaging of an NV ensemble

An alternative method to point-by-point imaging of the sample with a localized detector, is to create a spatially extended detector and record an image of the incident magnetic field at every point simultaneously. Here, a diamond wafer with a highly doped layer of NV centers is imaged onto a spatially resolving camera using wide-field microscopy (Fig. 5a). The PL intensity of each NV center is mapped to a particular pixel of the camera, allowing for parallel acquisition of the magnetic field amplitude over several hundred micron areas [Steinert2010,Pham2011,Chipaux2015,Simpson2016].

Wide-field imaging has been employed to image spin-labeled organic samples [Steinert2013], magnetotactic bacteria [LeSage2013], ferromagnetic thin films [Simpson2016], but also more exotic samples such as dusty olivine-bearing chondrules from the Semarkona meteorite [Fu2014]. As an illustration, a magnetic image of such a chondrule placed on a diamond imaging chip is shown in Fig. 5b. Here the measured magnetic field strength provides information about the nebular field that magnetized the chondrule in the first place [Fu2014].

In comparison to scanning techniques, wide-field microscopy typically suffers from much poorer spatial resolution, since diffraction limits the optical resolution to ~300 nm in most cases. The upside is that each pixel comprises the signal from a large number of NV centers ($N > 100$ typically), which leads to a gain in single-pixel sensitivity over a single NV magnetometer by a factor $\sqrt{N}$ [Rondin2014]. This, together with parallelized image acquisition, makes wide-field imaging ideal for applications where fast (video-rate, ~kHz) imaging of micrometer-sized samples is required. Improvements in spatial resolution down to a few nanometers can be obtained by applying super-resolution fluorescence microscopy techniques, but these are currently not compatible with parallelized acquisition [Rittweger2009,Han2010,Maurer2010,Pfender2014].

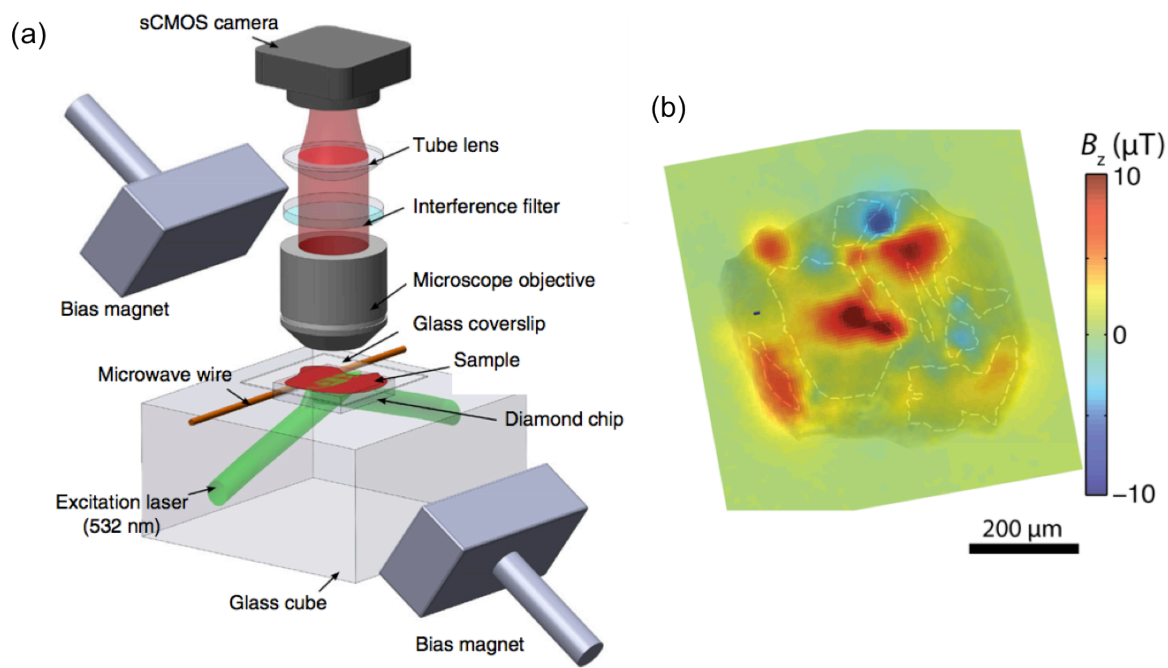

**Figure 5**: *(a) Principle of wide-field magnetic imaging. (b) Magnetic field map of an olivine-bearing chondrule from the Semarkona meteorite. Adapted from [Glenn2015] and [Fu2014].*

### 3.4 Challenges and further improvements

***Stand-off distance***. The two main figures of merit for NV-based magnetic imaging techniques are the spatial resolution and the magnetic sensitivity, which are ultimately set by the distance from the NV to the sample. Note that a smaller NV-sample distance also contributes to increasing the acquisition speed as it results in a larger magnetic field strength. Decreasing the NV-sample distance can be pursued in a variety of ways. Ion implantation can be used to deterministically define the NV distance to the surface within a few nanometers or less [Ofori-Okai2012]. The depth of nitrogen incorporation into the diamond is tuned by the implantation energy, and NV centers are created with a subsequent anneal at temperatures greater than 500°C. Recent techniques allow for implantation of individual atoms into diamond with nanometer spatial accuracy, or at the other end of the scale, the implantation dose can be increased by ten orders of magnitude to create high density ensembles of NV sensors. Using nanodiamonds is another way to constrain NV centers to be close to the diamond surface, where nanodiamonds as small as 20 nm have been used for NV magnetometry [Rondin2012]. A final method for obtaining near-surface NV centers is deterministic doping of the diamond during growth. Thin 'delta' layers of nitrogen containing diamond can be grown within a few nanometers of the diamond surface using chemical vapour deposition (CVD) growth of diamond [Mamin2013].

***Sensor readout***. As readout is performed by monitoring the NV defect PL intensity, the sensitivity of the NV magnetometer critically depends on the amount of PL signal that can be collected for processing, versus the amount of unwanted background light. Scanning confocal microscopy has advantages in that collection of background fluorescence is minimized, and highly efficient photodetectors can be used to convert nearly all (~80%) of the collected light into an electrical signal, whilst themselves producing little additional noise. In addition, fast detectors with nanosecond response times allow for temporal filtering of the fluorescence signal or investigations with high temporal resolution.

The diamond surface itself however, is a significant barrier to collection of the optical signal. Diamond's high refractive index prevents much of the PL signal from escaping the crystal and reaching the photodetector. To circumvent this issue, diamond fabrication techniques have been developed to enhance light collection. In particular, losses through total-internal-reflection can be minimized by shaping the diamond surface into a solid immersion lens [Marseglia2011,Robledo2011b], by using optical cavities [Jensen2014,Clevenson2015], or plasmonic and photonic structures enabling up to an order of magnitude increase in signal collection [Riedel2014, Choy2011, Momenzadeh2015]. However, effective integration of these optimized structures into magnetic imaging setups remains to be demonstrated. More suited to scanning probe experiments is the use of high aspect ratio pillars fabricated from bulk diamond [Babinec2011,Neu2014,Momenzadeh2015]: these structures act not only as waveguides to direct the NV emission towards the collection optics, but also as AFM scanning tips [Maletinsky2012].

There has additionally been much work towards improving the contrast of the ODMR signal beyond the 30% normally observed. By mapping the NV electronic spin state to the nuclear spin of the defect, multiple readout steps can be achieved, thereby enhancing the effective contrast [Steiner2010]. At moderate magnetic fields, the NV state can even be readout in a single shot [Jiang2009], bringing the dominant noise from photon shot noise into the quantum projection noise limit [Neumann2010]. Several other novel methods are also being explored to go beyond the photon noise limitation of PL detection. These include selective ionization of the NV defect depending on its spin state [Shields2015], and electrical readout of the photocurrent [Bourgeois2015].

***Diamond material***. Along with the placement of NV centers at the diamond surface, the diamond substrate itself can be optimized in order to enhance the sensitivity of the NV magnetometer. To improve sensitivity, the removal of extraneous noise is critical. Magnetic noise internal to the diamond arises from dopant spins, with the major species being substitutional nitrogen (a spin ½ defect known as the P1 center) and $^{13}$C nuclear spins. A single NV center in a diamond with 1.1% natural abundance $^{13}$C and part-per-million nitrogen content provides a magnetic sensitivity on the order of $0.1 - 1\ \mu T/\sqrt{Hz}$, which corresponds to detection of a single electron spin at more than 10 nm distance in one second of measurement time.

Ultrahigh purity CVD growth techniques can create diamonds with less than 1 ppb electronic spin and 10 ppb nuclear spin impurity content [Balasubramanian2009]. At this limit, phonon interactions with the diamond lattice become important, limiting the coherence time of the NV center to few milliseconds at room temperature. The magnetic sensitivity for a single NV center in this diamond reaches a limiting value of $\sim 1\ nT/\sqrt{Hz}$, which corresponds to the detection of a single proton at a distance of 10 nm in one second of measurement time. By cooling the diamond crystal, phonons are removed and the coherence time of the NV center increases. At cold temperatures, with high purity crystals, the detection of individual nuclei at tens of nanometer distance appears feasible.

However, NV centers close to the surface are sensitive not only to the bulk properties of the diamond substrate but also to surface defects and other surface-related effects. A major challenge is therefore to be able to place NV centers as close to the diamond surface as possible, while preserving the record sensitivities provided by deep NV centers in optimized ultrahigh purity diamond. To date, the magnetic sensitivities obtained from near-surface (< 5 nm) NV centers created with the above-mentioned techniques remain far inferior to state-of-the-art sensitivity. Several studies have evidenced the presence of magnetic noise caused by free spins on the surface [Romach2015,Myers2014,Rosskopf2014] but further work is required to fully understand the role of the surface and optimize the performance of NV magnetometers. In addition, shallow NVs suffer from photobleaching and fluorescence instability, the cause of which is also a subject of current research.

## 4. Applications

In this section, we illustrate how NV magnetometry can be used as a powerful tool (i) for the study of exotic spin textures in ferromagnets (§4.1) and (ii) for imaging single molecules (§4.2). We note that NV magnetometry finds other applications in a broad range of research domains, from paleomagnetism [Fu2014] to biomagnetism [LeSage2013], and electronic transport in condensed matter physics [Kolkowitz2015].

### 4.1 Imaging spin textures in ultrathin ferromagnets

Imaging magnetic fields with high sensitivity and nanoscale resolution is a key requirement for fundamental studies in nanomagnetism and the design of innovative ferromagnetic materials with tailored properties for applications in spintronics.

One particular class of ferromagnetic materials for which scanning-NV magnetometry appears particularly useful is ultrathin ferromagnets, where the ferromagnetic layer can be as thin as a few atomic planes (Fig. 6a). Domain walls (DWs) in such systems have attracted considerable interest over the last years owing to their potential use in low power spintronic devices. A domain wall (DW) is a region where the magnetization varies gradually between domains having different magnetization orientations. The DW width is in the range of 10-50 nm and is determined by the competition between the strength of the exchange interaction $A$, which tends to align spins with each other, and the magnetic anisotropy $K$, which orients spins in energetically favored directions. The possibility to move such nanoscale objects with currents is attracting great interest for the development of novel high-density memory storage devices. In particular, the observation of very large DW velocity (~400 m.s$^{-1}$ [Miron2011]) in ultrathin ferromagnets with perpendicular magnetic anisotropy (PMA) sandwiched between a heavy metal with large spin-orbit coupling and an oxide (Fig. 6a) promises groundbreaking opportunities for developing the racetrack memory protocol [Parkin2008]. Apart from these technological interests, current-induced DW motion along these asymmetric stacks have triggered *a major academic debate* regarding the underlying mechanisms of the DW motion [Ryu2013,Emori2013,Liu2012,Haazen2013]. Several models involving different spin-orbit torques, which could result from the Rashba effect and/or the spin-Hall effect, have been shown to be consistent with experimental observations [Martinez2013].

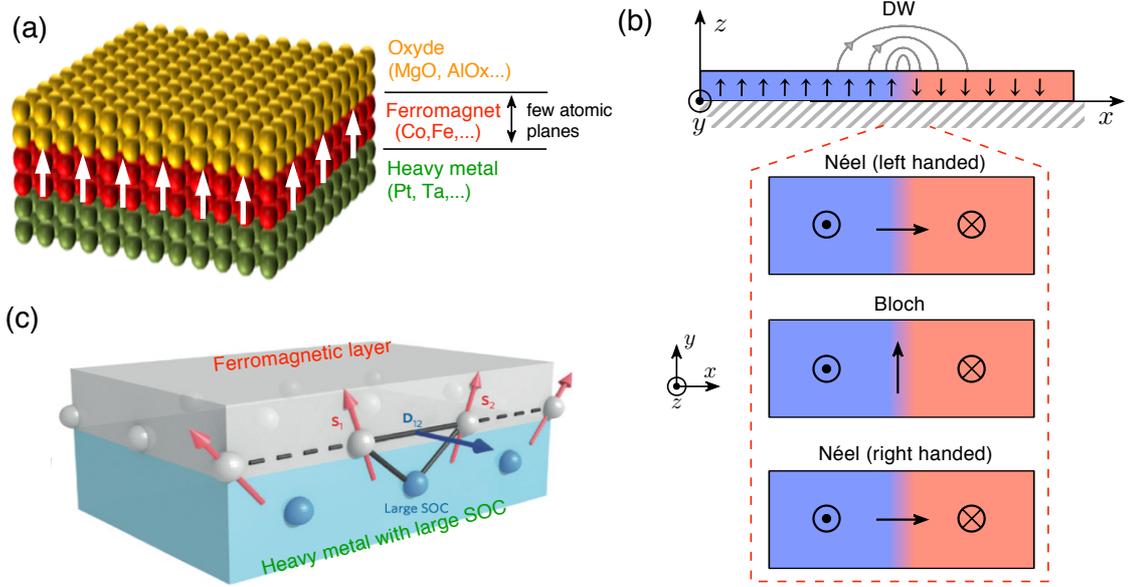

**Figure 6**: *(a) Ultrathin ferromagnetic layer with perpendicular magnetic anisotropy (PMA), sandwiched between a heavy metal with large spin-orbit coupling (SOC) and an oxide. (b) Schematic view of a DW. The black arrows indicate the internal magnetization while the grey arrows represent the magnetic field lines radiated above the film. The top views of the DW (lower panels) indicate the inner magnetization structure for Bloch, and Néel configurations. (c) Dzyaloshinskii-Moriya interaction (DMI) at the interface between a ferromagnetic layer and a heavy metal with large SOC (adapted from [Fert2013]).*

A key parameter of this debate is the nature of the DW, which can be of Bloch or Néel type (Fig. 6b), and dramatically affects the efficiency of the different proposed mechanisms. Indeed, the spin-orbit torque induced by the Rashba effect is efficient for a Bloch domain wall, while chiral domain walls of Néel type are efficiently driven by the spin Hall effect [Martinez2013]. In wide ultrathin wires with PMA, magnetostatics predicts that the Bloch DW, a helical reversal of the magnetization, is the most stable configuration since it minimizes demagnetization energy. Indeed, the magnetization in a Néel DW reverses in a cycloidal fashion, which adds an energy cost due to magnetic charges on each side of the wall. However, it was recently proposed that Néel DWs with fixed chirality could be stabilized by the Dzyaloshinskii-Moriya interaction (DMI), an indirect exchange occurring at the interface between a magnetic layer and a heavy metal substrate with large spin-orbit coupling [Thiaville2012]. The DMI between neighboring spins $S_1$ and $S_2$ induces an effective magnetic field given by $\overrightarrow{H_{DMI}} = \overrightarrow{D_{12}} \cdot (\overrightarrow{S_1} \times \overrightarrow{S_2})$, where $D_{12}$ is the Dzyaloshinskii-Moriya vector (Fig. 6c). Interface-induced DMI has been predicted from a three-site indirect exchange mechanism coupling two spins with a neighboring atom with large spin-orbit coupling [Fert1980], as shown in Fig. 6c. The DMI vector is then in the plane of the interface, and is expected to stabilize DWs in the Néel configuration, with a fixed chirality [Thiaville2012].

Direct, *in situ*, measurements of the inner structure of DWs in ultrathin ferromagnets would give a unique opportunity to quantify the strength of DMI and to clarify the physics of current-induced DW motion. This challenging task can be uniquely realized with scanning-NV magnetometry, through quantitative mapping of the stray field distribution generated above the DW [Tetienne2015]. This distribution is directly linked to the DW structure. Indeed, for a Bloch wall the only contribution to the stray field is the variation of the out-of-plane magnetization

$M_z$, while for a Néel wall, an additional contribution comes from the non-zero divergence of the in-plane magnetization $dM_x/dx \neq 0$, where $x$ is transverse to the DW (Fig. 6b). Analytical calculations show that the maximum relative difference of stray field between the Néel and Bloch configurations scales as $\pi\Delta_{DW}/2d$, where $\Delta_{DW}$ is the DW width and $d$ is the probe-to-sample distance. For typical ultrathin ferromagnets with PMA $\Delta_{DW}\sim 10$ nm, and the typical field values lie in the range of 1-5 mT at a distance $d=100$ nm. The difference of the stray field generated from a Bloch or a Néel configuration is therefore in the range of a few Gauss, which can be easily discriminated using scanning-NV magnetometry.

As an example, Fig. 7 shows typical results obtained on a Pt/Co(0.6nm)/AlO$_x$ trilayer system with PMA. The observation of current-induced DW motion with unexpectedly large velocities in this asymmetric stack was at the origin of the debate regarding the underlying mechanisms of the DW motion and their relationship with the DW structure. Magnetic field distribution recorded with scanning-NV magnetometry clearly shows a Néel-type DW structure with left-handed chirality, providing the first direct evidence of a strong DMI in a Pt/Co(0.6nm)/AlO$_x$ trilayer system [Tetienne2015]. Such a DW structure combined with spin orbit torques can explain all the characteristics of DW motion under current in this sample [Thiaville2012,Martinez2013]. We note that these results were then confirmed by measurements of DMI in similar samples using Brillouin Light Scattering (BLS) spectroscopy [Belmeguenai2015].

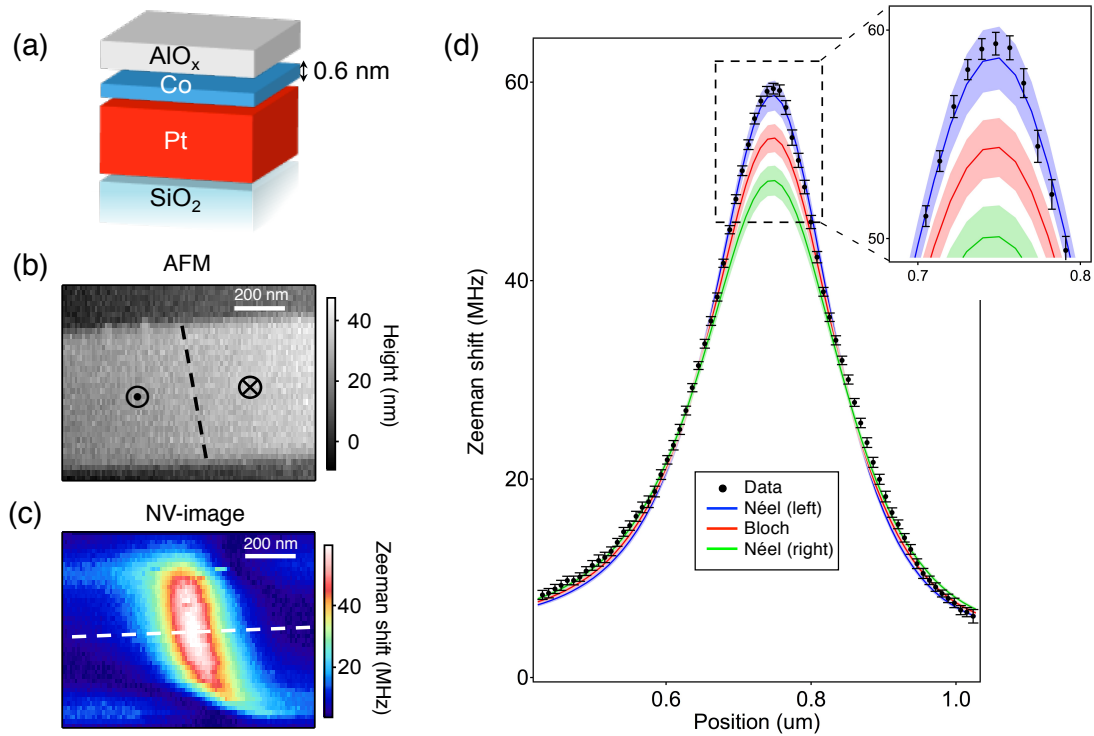

**Figure 7:** *(a) Schematic view of the ferromagnetic sample, a Pt(3nm)/Co(0.6 nm)/AlO$_x$(2nm) trilayer grown by sputtering on a SiO$_2$ wafer. (b) AFM image and (b) corresponding Zeeman shift map recorded by scanning the NV magnetometer above a DW in a 500-nm-wide magnetic wire of Pt/Co(0.6 nm)/AlOx. (d) Linecut extracted from (c) (see white dashed line), together with the theoretical predictions (solid lines). The shaded areas indicate the uncertainty in the simulations. Adapted from Ref. [Tetienne2015].*

Direct measurements of the DW structure in ultrathin ferromagnets can also be used to study the mechanisms at the origin of interfacial DMI, while tuning the properties of the magnetic material. To this end, the transition between Bloch and Néel DW driven by DMI can be analyzed by *(i)* changing the sample thickness, *(ii)* changing the ferromagnetic layer (Co, Fe, Ni) and *(iii)* by modifying the adjacent heavy metal underlayer (Ta, Pt) giving rise to strong spin-orbit coupling. In that spirit, it was recently shown that Bloch-type DWs are observed in a Ta/CoFeB(1 nm)/MgO trilayer stack, which indicates that interfacial DMI can be safely neglected at the Ta/$Co_{20}Fe_{60}B_{20}$ interface. By replacing Ta by W, the DMI is significantly increased, and is even strong enough to fully stabilize the DWs into the right-handed Néel configuration. This result shows that the direction of the DMI vector can be reversed by using W or Pt as heavy metal underlayer [Martinez2016]. Such studies might help to better understand the microscopic origin of interfacial DMI in ultrathin ferromagnets and to identify magnetic samples with large DMI strength that could sustain chiral spin textures.

Another striking phenomenon induced by DMI is the formation of exotic magnetic textures like spin helix, cycloids and skyrmion lattices. Magnetic skyrmions are chiral spin structures with a vortex-like configuration. As their structure cannot be continuously deformed to a ferromagnetic or other magnetic state, skyrmions are topologically protected, which makes them robust against material imperfections and thermal fluctuations. This remarkable topological protection enables skyrmion motion at ultralow current densities. Moreover, skyrmions can be as small as a few nanometers across and could potentially provide an ultrahigh information-storage density[1]. These properties make skyrmions attractive candidates for information storage and processing at the nanoscale [Fert2013].

Skyrmion lattices were first observed in 2009 in *bulk* non-centrosymmetric chiral magnets like MnSi, (Fe,Co)Si and FeGe (see [Nagaosa2013] for a recent review). In ultrathin films, skyrmion lattices of extremely small size have been observed very recently by spin-polarized scanning tunneling microscopy in a model sample consisting of one monolayer of Fe grown on Ir(111) [Heinze2011,Romming2013]. Despite these first promising results, significant challenges still need to be overcome before skyrmionic devices become reality. Indeed, a realistic device would require individual skyrmions to be nucleated and imaged in ultrathin ferromagnetic wires relevant for spintronic devices operating under ambient conditions. This challenge could be met in near future using scanning-NV magnetometry.

The experiments described in this section illustrate how scanning-NV magnetometry can be used as a powerful tool to study exotic spin textures in ferromagnetic nanostructures and to tackle fundamental problems in nanomagnetism. Importantly, the quantitative nature of the measurement can be directly exploited for direct comparison with micromagnetic simulations in order to discriminate between different theoretical models. We note that these methods can also be used to investigate spin wave excitations induced by ferromagnetic resonances [Wolfe2014,vanderSar2015], as well as the dynamics of thermally-activated Barkhausen jumps of the magnetic order [Tetienne2014,Dussaux2015]. Importantly all these experiments have been performed under ambient conditions, which is one of the key advantages of NV magnetometry. However, we note that low temperature (4K) operation was also recently reported, and will enable studying a new class of magnetic materials in the near future. As a proof-of-principle experiment, cryogenic NV-magnetometry was recently used to measure the stray field produced by superconducting vortices [Pelliccione2015,Thiel2015]. By placing the scanning-NV sensor in close proximity with the sample (<10 nm), the recorded magnetic field distribution reveals significant deviations from the widely used monopole approximation [Thiel2015]. In these experiments, it was shown that a more accurate description of the stray field distribution using Pearl's model is required in order to reproduce the experimental data

---

[1] More details can be found in the Chapter written by V. Cros.

(see Fig. 8). Once again, these experiments show how nanoscale and quantitative measurements provided by NV-magnetometry allow one to discriminate between competing theoretical models.

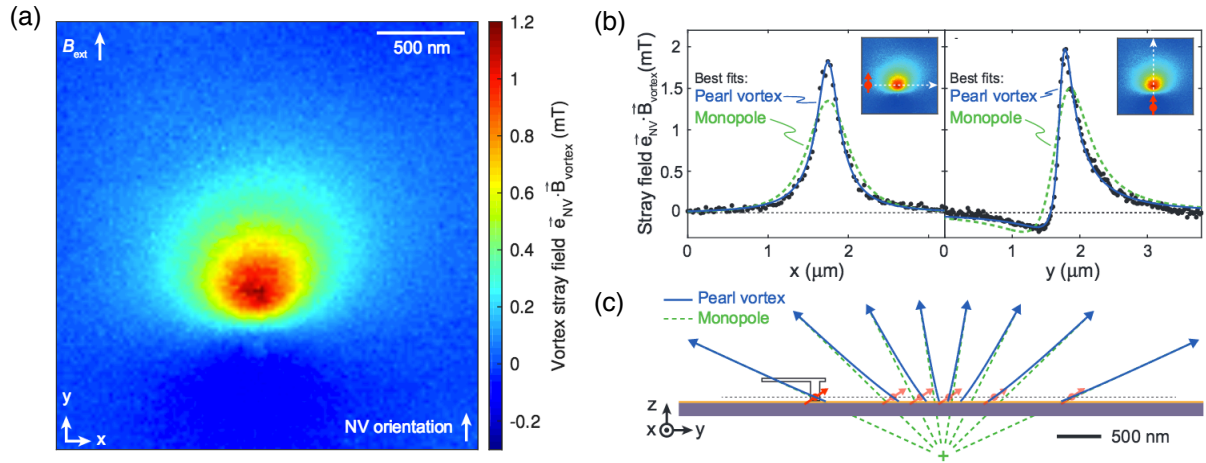

**Figure 8:** *(a) Quantitative magnetic imaging of the stray field above a superconducting vortex recorded with a scanning-NV magnetometer operating at 4 K. The width of the stray field distribution is fixed by the Pearl length (~800 nm). Here the distance between the scanning-NV sensor and the sample is around 10 nm. (b) Linecuts along the x and y axis extracted from (a) as shown in the insets. (c) Magnetic field lines using either the monopole or the Pearl vortex approximation. Adapted from Ref. [Thiel2015].*

## 4.2 Single molecule imaging and nanoMRI

Another promising application of diamond sensors capable of detecting sub-microtesla magnetic fields arising from nanometer sample volumes is in the field of nuclear magnetic resonance (NMR) spectroscopy. While NMR spectroscopy is amongst the most widely used methods for determining the structure of organic and biological molecules, the sensitivity is limited to macroscopic samples greater than tens of microns in dimension, meaning that detection is averaged over many millions of molecules. As a result, much of the microscopic information on sample heterogeneity – *e.g.* molecular sub-populations, individual configurations or orientations – is subsequently washed out by the ensemble measurement. The hurdles conventional NMR faces in detecting single molecules are the weak magnetic moment of nuclei and the difficulty to detect the microwave photons, which are emitted and absorbed by nuclei. In addition, typically low sample polarizations and the relatively weak coupling of the sample to microwave cavities hinder the sensitivity of inductive detection.

Diamond sensors are able to overcome these limitations by employing a simple principle: bringing the detector as close to the sample as possible. As the magnetic field of a single dipole reduces with distance $d$ as $1/d^3$, the magnetic field recorded by the sensor can be greatly enhanced by reducing the separation distance[2]. So much so, that the coupling between a single NV defect and sample spins can far exceed those achieved with high-Q microwave cavities, up to the point where the field from individual nuclei can be detected.

---

[2] For an unpolarized sample, the variance of the magnetic field is detected, which scales as $1/d^6$.

In 2013, the first nanoscale NMR spectroscopy experiments using NV defects ~ 5 nm from the diamond surface were performed [Staudacher2013, Mamin2013]. As shown in Fig. 9, the time-varying magnetic field arising from the precession of nuclear spins on the diamond surface is recorded by the NV defect. The magnitude of this detected magnetic field scales dramatically with the distance of the NV sensor to the nuclei, meaning that the detection volume is localized to ~ 100 nm$^3$ which corresponds to a few thousand molecules. The most widely used technique in these experiments is an extension of AC field detection described in §2.2 and is related to the Carr-Purcell-Meiboom-Gill sequence developed for NMR spectroscopy [Gullion1990]. A train of pulses is applied to the NV defect, each of which causes the NV spin to flip, and results in step-wise periods of forward and reverse phase accumulation. For a field whose sign changes in synchrony with the applied pulses, the NV spin acquires an overall phase due to this frequency filter centered about the flipping timescale (Fig. 9b). By adjusting the timing between the applied pulses, the frequency filter can be scanned to achieve a *lock-in* detection targeted at the nuclear Larmor frequency.

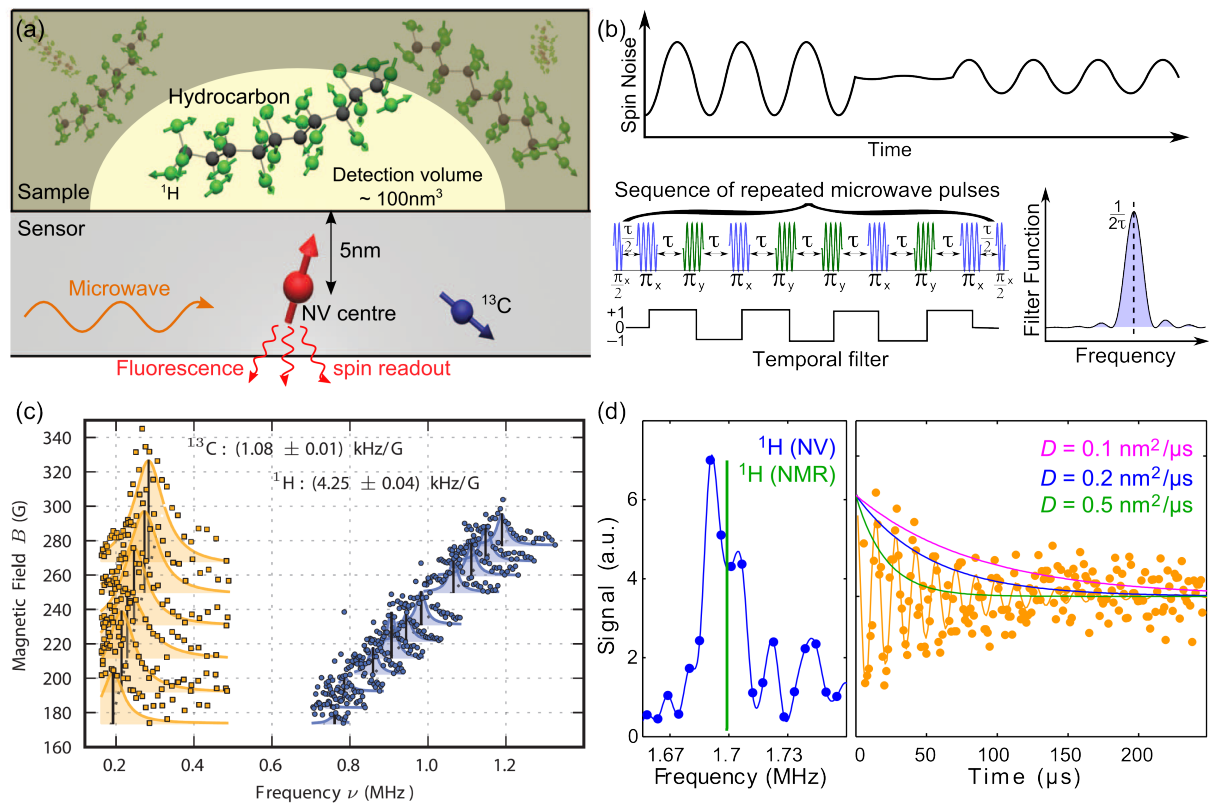

**Figure 9:** *(a) Nanoscale NMR spectroscopy of oil using an NV defect ~ 5nm from the diamond surface. (b) The sample magnetization is detected by applying a sequence of microwave pulses to the NV which creates a frequency filter that can be tuned to the Larmor frequency of target spins. (c) Example NMR spectra of protons (from oil) and $^{13}$C (from diamond) in ~100nm$^3$ detection volume. (d) Molecular diffusion broadens the NV-detected linewidth in comparison to the standard NMR linewidth. The broadening processes can be fitted to extract the sample diffusion rate, here ~0.2 nm$^2$/μs, for oil at the diamond surface. Adapted from Ref.* [Staudacher2013] *and* [Kong2015].

One immediate benefit to achieving nanoscale detection volumes is readily apparent from these initial experiments. The NMR spectra shown in Fig. 9c were obtained in external magnetic fields of a few hundred Gauss, meaning thermal polarization of the sample is negligible. Statistical polarization, however, is significant when the number of spins is of the order of $10^4$ or less. In spin ensembles there is a statistical probability that more spins are aligned in one direction at any point in time. The magnitude of the polarization overshoot from this excess alignment, is proportional to the square root of the number of spins in the ensemble, meaning that for small ensembles the statistical polarization can be a few percent, and reaches unity for single spins [Müller2014]. It is precisely these magnetic fluctuations in polarization around a zero mean value that are detected in the experiments shown in Fig. 9 and related work [Ohashi2013, Loretz2014]. The implication is that low or zero field NMR spectroscopy can be performed, thus significantly reducing the cost and complexity of experiments. Alternatively, by adjustment of the external static B field, the gyromagnetic ratio of sample spins can be identified (e.g. $^1$H, $^2$H, or $^{13}$C nuclei).

A second repercussion of nanoscale NMR is the ability to study molecular interactions with surfaces. Such investigations are not readily available to bulk techniques like conventional NMR spectroscopy where the surface area to volume ratio of samples is low. In particular, two studies have focused on the diffusion of liquids within a few nanometers of the diamond surface [Kong2015, Staudacher2015]. Using shallow NV defects, diffusion is observed as a broadening of the NMR spectral linewidth due to motion of molecules past the nanoscale detector (Fig. 9d). This line broadening effect is in stark contrast to liquid-state NMR where molecular diffusion decreases the spectral linewidth by reducing molecular interactions (an effect known as motional narrowing). By modeling the motion of molecules at the diamond surface, estimates of the molecular rotation and diffusion rates within few nanometer volumes can be obtained (Fig. 9d), giving information on boundary layer formation or molecular adhesion to the surface.

A further difference between conventional and nanoscale NMR spectroscopy is the spectral resolution. Initial studies with NV defects achieved several tens of kilohertz spectral resolution [Staudacher2013], which is many orders of magnitude worse than conventional NMR, and insufficient to resolve structural information such as chemical shifts. This drawback has meant that atomic level imaging has to-date, not been feasible. Recent developments however, have been able to boost the resolution of NV spectroscopy to a level now potentially capable of identifying different chemical groups [Kong2015, Staudacher2015, Laraoui2013].

The achievement of high-resolution spectroscopy coupled with single spin sensitivity [Müller2014, Sushkov2014b] promises the ability for structural analysis of single molecules, which has been a central driving force in the NV field. Yet while developments in nuclear spin spectroscopy are working towards this goal, NV sensors also find applications to detect molecules and nanoparticles with long-lived electron spins. High spin state electrons are of interest for their potential as contrast agents in magnetic resonance imaging, their vital role in cellular function, and the complex systems they form containing rich physics. However, just as with nuclear spins, very few techniques exist that are capable of detecting individual electron spins under ambient conditions.

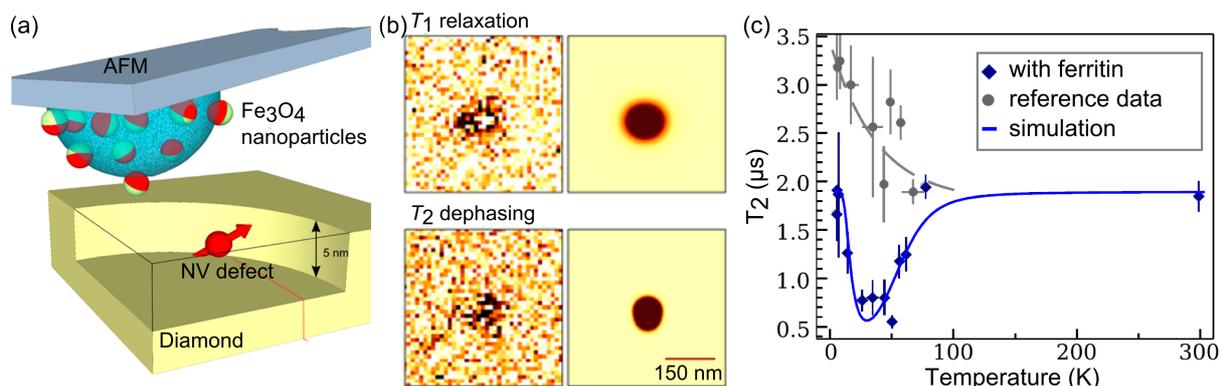

**Figure 10:** *(a) Magnetite nanoparticles are attached to an AFM tip and scanned over a shallow NV defect for magnetic imaging. (b) The NV defect operating in $T_1$ mode: sensitive to high (GHz) frequency fluctuations, and $T_2$ mode: sensitive to low (MHz) frequency fluctuations forms an image of a small ensemble of particles on the AFM tip. Shown on right are comparisons to simulations. (c) The NV defect operates across a temperature range of more than 300K, allowing magnetic phase transitions occurring in ferritin around 50K to be monitored. Adapted from Ref.* [SchmidLorch2015] *and* [SchäferNolte2015].

The NV center offers unique opportunities to non-invasively measure the dynamics of individual magnetic particles, with frequency bandwidth up to the gigahertz range. Scanning NV sensors are sensitive to the static component of the magnetic field generated by magnetic nanoparticles when operating in DC mode (see Fig. 2a), and operate with microtesla sensitivity and few nanometer spatial resolution. Access to fluctuating magnetic fields generated by magnetic particles can also be achieved by operating in AC (spin-echo) or relaxometry mode (see Fig. 2b,c). By attaching magnetite nanoparticles to an AFM tip, the magnetic field fluctuations of individual nanoparticles have been detected with NV centers (Fig. 10a). High frequency (gigahertz) fluctuations in the magnetization of magnetite particles induce spin relaxation of the NV center, whereas slow fluctuations induce dephasing of the NV center. In concert, these imaging modalities allow for imaging of the particle location with high resolution (Fig. 10b). Comparison of the signal strength in different frequency regimes allows for estimation of magnetic relaxation processes occurring within the nanoparticles and also the size of each nanoparticle [SchmidLorch2015].

The NV defect also has the advantage that it can operate over a wide temperature range, thus enabling investigations of temperature induced phase transitions in magnetic particles. As a result, the blocking temperature of molecular magnets can be explored or spin transitions in metalloproteins such as ferritin. Phase transitions are evidenced by changes in the magnetic field strength and fluctuation regime of ferritin molecules which in turn alter the dephasing rate of nearby NV sensors, thereby allowing temperature dependent spin relaxation dynamics of individual ferritin proteins to be unravelled [SchäferNolte2015] (Fig. 10c). In the biological setting, similar techniques can be applied to achieve highly sensitive detection of few molecules attached to nanodiamonds [Ermakova2013] and in solution [Ziem2013].

While the discussed experiments are mainly at the proof-of-principle stage, there is a great need for magnetic detectors in biology. The detection of electron spins in solution can be applied to the detection of free radicals, which play an important role in cell death and ageing. Furthermore, neural impulses generate a magnetic field, which although weak, could be imaged with an NV sensor. The application of NV centers to magnetic resonance imaging in biology may well see the tools of MRI, which we normally associate with millimeter scale resolution, now being used at the sub-cellular or molecular level.

# Conclusion

The field of nanomagnetism contains a wealth of opportunities for science and technology, which is being unlocked by a new generation of magnetic detectors. Amongst the magnetic sensors that are available to researchers today, the NV center in diamond stands alone in its ability to detect and image weak – nanotesla – magnetic fields with high – nanometer – spatial resolution at ambient conditions. This chapter has given a brief glimpse into some of the applications diamond sensors are finding.

The ability to image magnetic fields at the nanoscale will allow for improved characterization of magnetic storage devices and help investigation of new device concepts for information storage and processing. Nanoscale magnetometers will guide the development of new materials with new functionalities, not only ferromagnets but also superconductors or multiferroics. New sensors are intrinsically tied to the discovery of new physics, such as magnetic interactions at interfaces as outlined in section 4.1. Magnetic resonance at the nanoscale could allow for imaging of single molecules and has great potential for biology, as described in section 4.2. In the coming years, the list of applications of diamond sensors is expected to grow and will include not only nanomagnetism but many other fields of science. Diamond sensors have indeed great potential for thermometry and electrometry in addition to magnetometry, thus opening tremendous opportunities for nanocharacterisation.